\documentclass[letterpaper,epsf,pra,floatfix,amsmath,twocolumn,showpacs]{revtex4}

\usepackage{amsmath}
\usepackage{epsfig}
\usepackage{epsf}
\usepackage{array}
\usepackage{verbatim}
\usepackage{hyperref}

\newcommand{\re}[1]{(\ref{#1})}

\newcommand{\be}{\begin{equation}}
\newcommand{\ee}{\end{equation}}
\newcommand{\nn}{\boldsymbol{n}}
\newcommand{\rr}{\boldsymbol{r}}

\newcommand{\w}{\omega}
\newcommand{\rref}[1]{(\ref{#1})}
\newcommand{\eref}[1]{Eq.~(\ref{#1})}
\newcommand{\esref}[1]{Eqs.~(\ref{#1})}

\renewcommand{\Re}{\mathrm{Re}}
\renewcommand{\Im}{\mathrm{Im}}

\begin{document}

\title{Phase diagram, extended domain walls,  and soft collective modes in a three-component
fermionic superfluid }

\author{G. Catelani}
\author{E. A. Yuzbashyan}
\affiliation{Center for Materials Theory, Department of Physics and Astronomy,
Rutgers University, Piscataway, New Jersey 08854, USA}

\date{\today}

\begin{abstract}
We study the phase diagram of a three-component Fermi gas with weak
attractive interactions, which shows three superfluid and one
normal phases. At weak symmetry breaking between the components the
existence of domain walls interpolating between two superfluids
introduces a new   length scale much larger than the coherence
length of each superfluid. This, in particular,  limits the
applicability of the local density approximation in the trapped
case, which we also discuss.  In the same regime the system hosts
soft collective modes with a mass much smaller than the energy gaps
of individual superfluids. We derive their dispersion relations   at
zero and finite temperatures and demonstrate that their presence
leads to a significant enhancement of fluctuations near    the
superfluid-normal transitions.
\end{abstract}

\pacs{67.85.Fg, 67.85.Lm}

\maketitle

\section{Introduction}

In  Landau's approach to phase transitions, conventional superfluidity and superconductivity are characterized by a single complex order parameter. However,  in certain instances a proper  description of
a superfluid within this approach  requires the introduction of an order parameter with several complex components.
For example, in the case of superfluidity in $^3$He~\cite{hesup} the
spin-triplet, \textit{p}-wave pairing is described by nine  coefficients, and
three different superfluid phases are experimentally realized. Other
examples include unconventional superconductivity \cite{uscrev} in
heavy fermion compounds and color superconductivity in nuclear
matter~\cite{colsup}, where different phases could be realized
depending on the chemical potentials --  the two-flavor color
superconductor and the color-flavor-locked phase. Many aspects of
multicomponent superfluidity in these systems are understood only
on a phenomenological level due to their intrinsic complexity.
Atomic Fermi gases, on the other hand,  provide a unique
avenue to explore  these phenomena in a highly controllable way,
thanks to the tunability of the interactions between atoms.
Multicomponent superfluidity in atomic fermions could be realized,
for example, by trapping and cooling multiple hyperfine states of
the same atomic species \cite{liexp,lie2} or of different species
\cite{lkexp}.

Here we present a study of the phase diagram of a three-component Fermi gas
with weak attractive interactions.
In particular, we consider the situation in which the ``color'' symmetry
between the  components is   broken due to differences in the interaction strengths or
chemical potentials, while the masses are the same.
This situation is relevant to possible experiments involving three
hyperfine states of $^6$Li atoms. We first develop a Ginzburg-Landau
expansion for this system and use it to confirm   the previous
results \cite{hof,tor1,demler} that there are four possible phases:
the normal state and three superfluid states S$_1$, S$_2$, and
S$_3$. In each of the superfluid states two out of three components
are paired, while the third one is in a normal state. First order
phase transitions  between different superfluid states  can be
driven by varying interaction strengths, chemical potentials,
particle densities  or temperature.  We construct the phase diagram
in grand canonical and canonical ensembles at finite and zero
temperatures, shown in Figs.~\ref{fig:1}-
\ref{fig:g3},  and \ref{fig:4}; see also
Refs.~[\onlinecite{tor1,demler}].

The canonical phase diagram at fixed temperature (Fig.~\ref{fig:1b})
contains regions where the homogeneous state is unstable. When the
particle densities are within these regions the two superfluid
states phase separate, as expected for the first-order phase
transition; see, e.g., \cite{LL5}. We therefore explore the
properties of domain walls between different superfluids; see
Fig.~\ref{fig:2}. In particular, we explicitly determine the shape
and the thickness $\ell$ of the domain walls in various regimes. The
case of weak symmetry breaking between two components (say, 1 and
2) -- i.e., when   the couplings of 1 and 2 with 3 and chemical
potentials $\mu_1$ and $\mu_2$ are close -- is especially interesting.
 At full symmetry between 1 and 2, $\ell=\infty$. This is natural as in this case the
thermodynamic potential $\Omega(\Delta_1,\Delta_2)$, where  $\Delta_1$ and $\Delta_2$ are the order parameters
for superfluid states S$_1$ and S$_2$, respectively, is invariant with respect to rotations in the $\Delta_1$-$\Delta_2$ space.
This implies that the two minima of the potential $(\Delta_1,\Delta_2)=\{(\Delta^0_1,0), ( 0,\Delta^0_2)\}$ describing superfluids
 S$_1$ and S$_2$ can be connected by continuous lines of minima.
Then, $\Delta_1$ can be continuously  deformed  into $\Delta_2$ at no energy cost when moving from one point in space to another.
At weak symmetry breaking, as we demonstrate below, the thickness of the domain wall $\ell$ is parametrically larger than the coherence
lengths $\xi_1$ and $\xi_2$ of superfluids S$_1$ and S$_2$.

For a trapped three-component gas the
local density approximation (LDA) predicts sharp boundaries between superfluid
states S$_1$ and S$_2$ \cite{tor2}. In reality, there is a domain wall of length $\ell$
between S$_1$ and S$_2$ where the two superfluids coexist. Therefore,
the characteristic length scale over which the boundaries predicted by the LDA are smeared is
$\ell$, rather than $\xi_1$ or $\xi_2$. Moreover, if the radius of the trap, $R$, is comparable to
$\ell$, the two superfluids coexist throughout the trap, so that the cases $R\gg\ell$ and
$R\approx \ell$ are qualitatively different. In particular, this means that
the LDA breaks down in the entire trap when $R\lesssim \ell$. For typical experimental parameters, the
condition $R\gg\ell$ translates into $N_t\gg 10^4$ (see below), where $N_t$ is the total  number of
fermions of all three species. Moreover, as we will
see, the deviations from the LDA are significant already for $N_t$ as large as $10^7$.

Another consequence of the weak symmetry breaking discussed above
is the presence of soft collective modes in multicomponent Fermi gases \cite{hof,ssb}. Suppose, for example, the system is in the superfluid state S$_1$. By considering  the
thermodynamic potential $\Omega(\Delta_1,\Delta_2)$ as above in the case of the domain wall, we expect  fluctuations
 $\delta \Delta_2(\rr, t)$ towards superfluid S$_2$ to be massless in the symmetric case. Below  we derive the mass
and dispersion relations of the corresponding collective  modes in the general asymmetric case at $T=0$ and at finite temperatures.
We show that at weak symmetry breaking the mass can be much smaller than the   BCS gap in the superfluid S$_1$ in the absence of
the third fermionic species. At finite temperature these fluctuations result, in particular, in an enhancement of
the Ginzburg-Levanyuk number $Gi$ by a large factor. In other words, the window of temperatures around the critical temperature
for the normal - superfluid S$_1$ transition where fluctuations dominate becomes much larger in the presence of the third
component.

Let us comment on the experimental realization of superfluidity in a three-component Fermi gas.
Achieving a stable gas in this case appears more challenging than in a two-component one due to
the enhanced role of  the three-body scattering. In the two-component Fermi gas three-body recombination is suppressed
thanks to the Pauli exclusion principle \cite{shly} and the system is stable over tens of seconds.
In the three-component case there is no such suppression and the decay time is of the order
of milliseconds \cite{lie2}. Various proposals are being put forward in order to increase the lifetime
of the system, such as, e.g., the stabilization by an optical lattice \cite{gurarie} similar to that for bosonic atoms \cite{boslat}. We note that the results presented in this paper are obtained
in the weak-coupling regime, which is expected to be insensitive to the stabilization technique.
For example, a lattice added to the trapping potential affects the single-fermion spectrum only.
This is irrelevant at weak coupling since the superfluid energy scales are assumed to be much smaller than the fermionic
bandwidth. The single-particle bands contribute only through the density of states at the Fermi energy
irrespective of the details of the spectrum.

The paper is organized as follows. In the next section we give a brief overview of
the mean-field approach and introduce our notation. In Sec.~\ref{glexp} we present   the
Ginzburg-Landau expansion of the thermodynamic potential and discuss the phase diagram at
finite temperatures. We study the domain walls in Sec.~\ref{sec:dw}, and
in Sec.~\ref{ztpd} we describe the zero-temperature phase diagram.
Section~\ref{sec:cm} is devoted to the collective modes. Finally, we summarize
our results in Sec.~\ref{summ}.

\section{Thermodynamic potential}
\label{tpc}

In this section we outline the derivation of the thermodynamic potential, from
which the phase diagram and all thermodynamic quantities can be obtained.
We will not go into details, as the derivation is a well-known procedure \cite{textb}.
Our starting point is the following Hamiltonian:
\be
H=\sum_{i=1}^3 \psi_i^\dag H_0 \psi_i + H_{\mathrm{int}} \,.
\ee
Here $H_0={\bf p}^2/(2m)$ is the single-particle Hamiltonian (we assume that all the particles have the
same mass). As discussed in the Introduction, an optical lattice would modify the single-particle
Hamiltonian. Its effect can be taken into account by introducing an effective mass
$m_\mathrm{eff} \neq m$, which in the weak-coupling regime results
only in a renormalization of the density of states introduced below in \eref{GLe}.
The pairwise interaction part is
\be
H_{\mathrm{int}} = \sum_{i,j,k,j',k'} \frac{g_i}{4} \left(\psi_j^\dag \varepsilon_{ijk} \psi_k^\dag\right)
\left(\psi_{k'} \varepsilon_{ij'k'} \psi_{j'}\right),
\ee
where $\varepsilon_{ijk}$ is the totally antisymmetric tensor and $\{i,j,k\}=\{1,2,3\}$.
By the Hubbard-Stratonovich transformation, we introduce the pairing field $\vec{\Delta}(\tau, \rr)=\left( \Delta_1(\tau, \rr), \Delta_2(\tau, \rr),\Delta_3(\tau, \rr)\right)$
and after integrating out the particle fields $\psi_i$, we obtain the following effective action
for $\vec{\Delta}(\tau, \rr)$:
\be\label{effac}
S_{\mathrm{eff}} \left\{\vec{\Delta}\right\} = \int \!d\tau \, d^3x
\left[\vec{\Delta}^\dag \hat{g^{-1}}\vec{\Delta}
-\frac{1}{2} \ln \mathrm{det} \hat{G}^{-1}\right],
\ee
where $\hat{g^{-1}}=\mathrm{diag}\left( g_i^{-1} \right)$ and $\hat{G}^{-1}$ is the particles' inverse
Green's function, which is a $6\times 6$ matrix in Nambu-Gorkov space with the structure
\be
\hat{G}^{-1} =
\left(
\begin{array}{cc}
\left(-\partial_\tau - H_0 +\mu_i \right) \delta_{ij} &
\epsilon_{ijk} \Delta_k \\
-\epsilon_{ijk} \Delta^\dag_k &
\left(-\partial_\tau + H_0 -\mu_i \right) \delta_{ij}
\end{array}
\right).
\ee
Here $\mu_i$ are the chemical potentials for the different species.

In the mean-field approximation the thermodynamic potential is obtained by evaluating
the effective action (\ref{effac}) for a $\tau$-independent pairing field $\vec\Delta(\rr)$.
This is expected to be an excellent approximation for the description of a weakly coupled
fermionic superfluid at temperatures not extremely close to the transition temperature
\cite{LL5}.
First, let us consider the case of a uniform order parameter. Performing
a Fourier transform from real space-imaginary time
to the momentum--Matsubara-frequency space in \eref{effac} we derive
\be\label{tpeq}\begin{split}
\Omega = & -\sum_i \frac{|\Delta_i|^2}{g_i}+\int\!\frac{d^3 p}{(2\pi)^3} \,
\bigg\{\sum_i \frac12 \xi_i \\
& - \frac{1}{2\beta}\sum_{n} \ln \bigg[-2\prod_i
\left(\w_n^2+\xi_i^2\right) \\ & +\sum_P \left(\w_n^2+\xi_i^2\right)
\Big|\left(\w_n+\mathrm{i}\xi_j\right)\left(\w_n-\mathrm{i}\xi_k\right)+|\Delta_i|^2\Big|^2
\\ &
+\sum_P |\Delta_i|^2|\Delta_j|^2\Big[\left(\w_n+\mathrm{i}\xi_i\right)\left(\w_n-\mathrm{i}\xi_j\right)
+ \mathrm{c.c.}\Big]
\bigg] \bigg\} ,
\end{split}
\ee
where $\omega_n = 2\pi T (n+1/2)$, $\xi_i={\bf p}^2/(2m)-\mu_i$, the sum over $P$ denotes the sum over cyclic permutations
of $\{i,j,k\}=\{1,2,3\}$, and ``c.c.'' is the complex conjugate.
For vanishing order parameter $\vec{\Delta}=0$, we obtain
the sum of the thermodynamic potentials for three perfect gases, as expected. Also, for an order
parameter with only one nonvanishing component $\Delta_i\neq 0$, \eref{tpeq} reduces to the sum
of the potentials of a normal gas and a two-component Fermi superfluid. Let us denote
 the corresponding zero-temperature order parameter of the two-component superfluid in the absence of the third
 fermionic species as $\Delta_i^0$.
We note that \eref{tpeq} is
ultraviolet divergent, and a regularization procedure (e.g., a hard cutoff as for superconductors
\cite{LL5} or a $T$-matrix approach \cite{review})
should be implemented. Then all physical quantities can be expressed in terms of
the $\Delta_i^0$'s, as we do in what follows.

The (meta)stable states are given by the (local) minima of $\Omega$. This condition
determines the mean-field phase diagram. We will show below that the possible phases
fall into two classes -- normal state or a two component superfluid
plus a normal gas -- in agreement with the results of \cite{tor1,demler}.
The two superfluid components can be any two of the three atomic species; i.e., there
are three possible superfluid states, which we denote as S$_1$, S$_2$, and S$_3$ when the paired species are
2 and 3, 1 and 3, and 1 and 2, respectively.
For simplicity, unless otherwise specified, we  assume from now on $g_3=0$, so that $\Delta_3 = 0$ and only $\Delta_1$ and $\Delta_2$
components of the order parameter can be nonzero. This can
be a good approximation in the case of three hyperfine states of $^6$Li, where two
out of three Feshbach resonances
mediating the attractive interactions between the states are close in magnetic
fields \cite{liexp}. The third
resonance is at a lower field and can be neglected on the BCS side of the crossover.
The inclusion of the case $g_3 \neq 0$ in our formalism is straightforward. We briefly comment on this
case in Sec.~\ref{g3em} and show the corresponding phase diagram
in Fig.~\ref{fig:g3}.
For concreteness, we take $|g_1| > |g_2|$ and introduce the notation:
\be\label{difem}
h_1=\mu_3-\mu_2 \, , \quad
h_2=\mu_3-\mu_1 \, ,
\ee
for the differences in chemical potentials.

\section{Ginzburg-Landau expansion}
\label{glexp}

Here we perform a Ginzburg-Landau expansion for the thermodynamic potential and use
it to obtain the finite-temperature phase diagram of the system in the $h_1$-$h_2$ plane;
see Fig.~\ref{fig:1}.
We determine the superfluid-superfluid and superfluid-normal transition lines and the
metastability regions in both grand-canonical (Fig.~\ref{fig:1}) and
canonical (Fig.~\ref{fig:1b}) ensembles. In the latter case there is a region of the
phase diagram where a homogeneous state is unstable and a phase separation between two
types of superfluid takes place. We identify this region as well as the corresponding
supercooling lines; see Fig.~\ref{fig:1b}.

According to Landau's phenomenological approach \cite{LL5}, the thermodynamic potential near
a second-order phase transition can be expanded in powers of the order
parameter. If only even powers are present and the coefficient
of the fourth-order term is positive, the vanishing of the coefficient of the
quadratic term determines the second-order transition point. When the fourth-order term also changes sign,
the transition becomes first order, and higher-order terms should be included in the power series.

As shown by Gorkov \cite{gork}, this phenomenological theory can be  derived
by expanding the microscopic theory   in $|\vec\Delta|/2\pi T$ around $\vec{\Delta}=0$.
Using \eref{effac} [or \eref{tpeq} for the uniform part], we obtain  to the fourth order in components of $\vec{\Delta}=(\Delta_1,\Delta_2,0)$
\be\label{GLe}\begin{split}
\Omega - \Omega_N  = & \nu \sum_{i=1}^2\left\{ \alpha_i |\Delta_i|^2 +\frac{\beta_i}{2}\left[|\Delta_i|^4 +
\frac{v_F^2}{3}|\nabla \Delta_i|^2\right]\right\} \\
& +\nu\beta_{12} |\Delta_1|^2|\Delta_2|^2,
\end{split}\ee
where $\Omega_N$ is the normal-state thermodynamic potential   of the ideal gas, $\nu$ is the density
of states at the Fermi energy, and
  the coefficients $\alpha_i$, $\beta_i$, and
$\beta_{12}$  are
\be\label{alphaiem}
\alpha_i=\ln\frac{T}{T_{c_i}} + \Re\Psi\left(\frac{1}{2} +   \mathrm{i}\frac{h_i}{4\pi T}\right)
- \Psi\left( \frac{1}{2}\right),
\ee
\be\label{betaiem}
\beta_i = -\frac{1}{4}\frac{1}{(2\pi T)^2}\Re\Psi''\left(\frac{1}{2}+\mathrm{i}\frac{h_i}{4\pi T}\right),
\ee
and
\be\label{b12em}\begin{split}
\beta_{12} =\frac{1}{h_1-h_2} \frac{1}{4\pi T} &
\Im \bigg[\Psi'\left(\frac{1}{2}+\mathrm{i}\frac{h_2}{4\pi T}\right)
\\ & \qquad
 -
\Psi'\left(\frac{1}{2}+\mathrm{i}\frac{h_1}{4\pi T}\right)\bigg].
\end{split}\ee
Here $\Psi(x)$ is the digamma function and $T_{c_i}$ is the critical temperature of the superfluid S$_i$ at zero
chemical potential difference and in the absence
of the third fermionic species -- i.e., for $h_i=0$ and $g_{j\ne i}=0$. According to the standard BCS theory
for two species, $T_{c_i}$  is related to the corresponding zero-temperature
order parameter $\Delta_i^0$ as $T_{c_i}=e^{\gamma_E}  \Delta_i^0/\pi$, where $\gamma_E$ is Euler's constant.
Note that due to our choice $|g_1|>|g_2|$ for the coupling constants, $T_{c_1}>T_{c_2}$.
We will comment below on the physical meaning of the temperatures $T_{c_i}$ in the  three species case.

Expressions \re{alphaiem}, \re{betaiem}, and \re{b12em} for the coefficients in
the Ginzburg-Landau expansion (\ref{GLe}) were derived in the weak-coupling limit, which enabled us to approximate
the density of states with a constant $\nu$. However,
the structure of the potential \re{GLe} is dictated by symmetry and must remain the same  at any coupling. Indeed, since the particle
number is conserved separately
 for each species, the potential must be independent of
the phases of the complex components of the order parameter. Therefore, the only allowed
terms in the expansion  to the fourth order are $|\Delta_i|^2$, $|\Delta_i|^4$, and
$|\Delta_1|^2|\Delta_2|^2$.

The terms in curly brackets in \eref{GLe} give the thermodynamic potential $\Omega_i(\Delta_i, h_i,T)$ of the two component superfluid $S_i$ in the
absence  of the third species, while the  $|\Delta_1|^2|\Delta_2|^2$ term represents the interaction
between the two superfluids. The same
expression for $\Omega_i(\Delta_i, h_i,T)$ was previously obtained \cite{BK}
in a study of the nonuniform superconducting
Fulde-Ferrell-Larkin-Ovchinnikov (FFLO) state \cite{FFLO}. In thin superconducting
films in a parallel magnetic field the
 thermodynamic potential $\Omega_i(\Delta_i, h_i,T)$   describes the effect of the Zeeman splitting. In this case,
$h_i$ has a meaning of the Zeeman magnetic field and $\Delta_i$ is the superconducting order parameter. Let us briefly summarize the phases described by $\Omega_i$ in the $h_i$-$T$ plane \cite{fulde} before we
proceed to the phase diagram for three species. For $T>T_{c_i}$ the quadratic coefficient is positive, $\alpha_i>0$, and
the two-component Fermi gas is in the normal state for any value of $h_i$. At  temperatures $T^i_{\mathrm{tri}}<T<T_{c_i}$
a second-order transition to the superfluid state S$_i$ takes place when $\alpha_i(h_i, T)=0$. For temperatures lower than the
tricritical temperature,
\be\label{triem}
T^i_{\mathrm{tri}}\simeq 0.56 T_{c_i},\quad i=1,2,
\ee
 the quartic coefficient  is negative, $\beta_i<0$, whenever $\alpha_i\to 0$
and the normal-superfluid S$_i$ transition is first order. The tricritical temperature and the corresponding tricritical chemical potential are determined from the condition  $\alpha_i(h^i_{\mathrm{tri}},T^i_{\mathrm{tri}} )=\beta_i(h^i_{\mathrm{tri}}, T^i_{\mathrm{tri}})=0$. This picture can also be obtained in the BCS limit from the phase diagram for polarized Fermi
gases in the BCS-BEC crossover \cite{rad,parish}.

Now we turn to the analysis of general properties of the full thermodynamic potential  for three species.
The Ginzburg-Landau expansion \re{GLe} is a good approximation only when  the polynomial $\beta_1 |\Delta_1|^4+\beta_2 |\Delta_2|^4 + 2\beta_{12}|\Delta_1|^2|\Delta_2|^2$
is positively defined. Otherwise, $\Omega\to-\infty$ as $|\vec\Delta|\to\infty$ along a certain direction
in the $\Delta_1$-$\Delta_2$ plane. Using Eqs.~\re{betaiem} and \re{b12em}, one can show that this condition
reduces to
\be\label{bidef}
\beta_i = -\frac{1}{4}\frac{1}{(2\pi T)^2}\Re\Psi''\left(\frac{1}{2}+\mathrm{i}\frac{h_i}{4\pi T}\right) > 0, \quad i=1,2.
\ee
These inequalities are equivalent to $|h_i|/4\pi T\lesssim 0.304$. Since Eq.~\re{GLe} was obtained by expanding in $|\vec \Delta|/2 \pi T$, we
also should have $|\vec \Delta|/2\pi T\ll 1$. Thus, the conditions of applicability of the expression \re{GLe}
for the thermodynamic potential are
\be\label{GLcond}
\frac{|h_i|}{4\pi T}\lesssim 0.304,\quad \frac{|\vec \Delta| }{2\pi T}\ll 1, \quad i=1,2.
\ee
It follows from the discussion in the previous paragraph that these inequalities  hold only
in the ``high-temperature'' regime $T> T^1_{\mathrm{tri}}>T^2_{\mathrm{tri}}$. Otherwise, $\beta_1<0$ when $\Delta_1/2\pi T\ll 1$
and the first condition in Eq.~\re{GLcond} is violated.
For the remainder of this section we restrict ourselves to this range of temperatures. Then, one can show using Eqs.~\re{alphaiem}
and \re{betaiem} that the condition \re{bidef} always holds whenever any of the quadratic coefficients $\alpha_i$ is sufficiently
small.

Let us discuss the possible phases of the three species system in the $h_1$-$h_2$ plane as a function of temperature
going from higher to lower temperatures. For $T>T_{c_1}>T_{c_2}$ we see from Eq.~\re{alphaiem} that both
quadratic coefficients in Eq.~\re{GLe} are positive, i.e.,
\be\label{nsstab}
\alpha_i=\ln\frac{T}{T_{c_i}} + \Re\Psi\left(\frac{1}{2} +\mathrm{i}\frac{h_i}{4\pi T}\right)
- \Psi\left( \frac{1}{2}\right) > 0 \, ,
\ee
for any $h_i$ and $i=1,2$. In this case, the only stable state is $\vec{\Delta}=0$ -- i.e., the normal state.
As the temperature is lowered,  $\alpha_1$ first vanishes at $T=T_{c_1}$ and $h_1=0$, while $\alpha_2$ remains
positive. Therefore, a phase transition from the normal to the superfluid state
$S_1$ occurs and $T_{c_1}$ is
the  actual critical temperature for this transition. Generally,
for $T_{c_2}<T<T_{c_1}$, we have $\alpha_2>0$, while $\alpha_1$ changes sign at a
temperature-dependent critical chemical potential $h_1^c(T)$
determined by the equation $\alpha_1\left(h_1^c(T),T\right)=0$. At $h_1<h_1^c(T)$ the superfluid
state $S_1$ is the stable one, while at larger $h_1$ the system turns normal.

The case $T<T_{c_2}$ is more complicated. Now the conditions \re{nsstab}
hold for both components only when both $|h_1|$ and $|h_2|$ are sufficiently large.
In the $h_1$-$h_2$ plane Eq.~\re{nsstab} determines four normal-state regions; see Fig.~\ref{fig:1}.
A second-order phase transition from the normal to a superfluid
state takes place when one of the coefficients $\alpha_i$ changes sign. For example, starting from
the normal state, keeping $h_2$ fixed, and changing $h_1$, we get  a transition between the normal state and the
superfluid $S_1$, as $\alpha_1$ becomes negative while $\alpha_2$ is still positive. This argument,
however, cannot predict the state of the system in the central region of the $h_1$-$h_2$ plane where
both $\alpha_i$'s are negative. We will explore this region in more detail in the following subsection.

\subsection{Phase diagram in the vicinity  of critical temperatures}
\label{sec:pdtc}

\begin{figure}
\begin{flushleft}\includegraphics[width=0.45\textwidth]{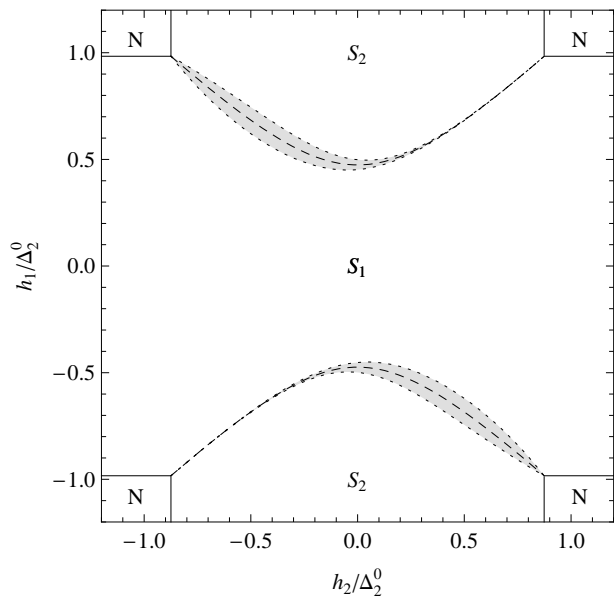}\end{flushleft}
\caption{Finite-temperature ($T=0.85T_{c_2}$) phase diagram for a three-component Fermi gas in
the $h_1$-$h_2$ plane of chemical potential differences, \eref{difem}.
The two nonvanishing pairwise couplings $g_{1,2}$ are such that $T_{c_1}/T_{c_2}=1.04$,
where $T_{c_i}$ are defined below \eref{b12em}.
Note that the normal state (N) is stable at
large $h_i$ while the superfluid states (S$_i$) at lower ones.
The stronger interaction ($|g_1|>|g_2|$) determines the superfluid state (S$_1$) realized at
$h_1=h_2=0$.
The horizontal (vertical) segments denote second-order N-S$_1$ (N-S$_2$) transitions.
The dashed curves mark the first-order S$_1$-S$_2$ transitions; see \eref{degcon}.
The shaded areas limited by the dotted curves [\eref{stabcond}] are the metastability regions.}
\label{fig:1}
\end{figure}

As discussed above, the normal state is the stable phase
in four sectors of the phase diagram, corresponding to the four corners in Fig.~\ref{fig:1}.
Here we show that in the central region two different
cases are possible:
(i) the thermodynamic potential has only one minimum, which coincides with the superfluid state S$_i$ for
 one of the two possible condensates $\Delta_i$; (ii) $\Omega$ has two local minima, such that one
condensate is
the stable state and the other one is a metastable one. In the latter situation, a first-order phase
transition separates the two superfluid states, as identified by the dashed lines in Fig.~\ref{fig:1}.  The two minima are degenerate along these lines in the $h_1$-$h_2$ plane. The gray areas
around the lines shown in Fig.~\ref{fig:1} enclose the regions where two local minima are present.

In this subsection we obtain the phase diagram for the case when the
two coupling constants $g_1$ and $g_2$ are sufficiently close in
magnitude.  We also take the
temperature to be near the critical temperatures $T_{c_2}$ and  $T_{c_1}$, i.e.,
\be\label{closeem}
T_{c_1}-T_{c_2} \ll T_{c_2},\quad T_{c_2}-T\ll T_{c_2}.
\ee
The first inequality in Eq.~\re{closeem} holds since  $T_{c_i}$ is the critical temperature for
the two component superfluid with coupling $g_i$ [see the text below Eq.~\re{b12em}] and the
couplings are close. As we will see below, in this case the
condition $|\vec\Delta|/2\pi T \ll 1$ for the validity of the
Ginzburg-Landau expansion holds. Then, it
follows from Eq.~(\ref{GLcond}) that expression (\ref{GLe}) for
the thermodynamic potential can be used not just near the phase
transition lines, but for all $h_1$ and $h_2$ such that $|h_i|/4\pi T\lesssim 0.304$.
Nevertheless, the conclusions we  draw regarding the phase diagram have
general validity at sufficiently high temperatures   $T> T^1_{\mathrm{tri}}\simeq 0.56 T_{c_1}$;
see the text below Eq.~\re{GLcond} and at the end of this subsection.

Let us first consider a homogenous system; i.e., the gradient terms in \eref{GLe} vanish:
\be\label{tpho}\begin{split}
& \Omega \left(|\Delta_1|^2,|\Delta_2|^2\right)  - \Omega_N = \\
& \nu \sum_{i=1}^2\left\{ \alpha_i |\Delta_i|^2 +\frac{\beta_i}{2}|\Delta_i|^4 \right\}
+\nu\beta_{12} |\Delta_1|^2|\Delta_2|^2.
\end{split}\ee
To find the stationary points of
$\Omega$, we pass to a polar coordinate representation
\be\label{poldec}
|\Delta_1|= \Delta \cos\theta \, , \quad |\Delta_2|=\Delta\sin\theta \, .
\ee
Differentiating \eref{tpho} with respect to the angular variable $\theta$, we find
\be\label{theom}\begin{split}
& 0 = \Delta^2 \cos\theta\sin\theta\Big[\alpha_2  - \alpha_1
\\ & + \Delta^2 \left(\left(\beta_{12}-\beta_1\right)\cos^2\theta
+\left(\beta_{2}-\beta_{12}\right)\sin^2\theta \right)\Big],
\end{split}\ee
where $\alpha_i$, $\beta_i$, and $\beta_{12}$ are defined by Eqs.~(\ref{alphaiem}),
(\ref{betaiem}), and \re{b12em}, respectively.
Equation~(\ref{theom}) always admits the three solutions $\Delta=0$, $\theta=0$, and $\theta=\pi/2$.
These are, respectively, the normal state, the condensate $\Delta_1$, and the condensate
$\Delta_2$. To determine the value of the nonvanishing order parameter component, we also need
to equate to zero the derivative of the thermodynamic potential \rref{tpho} with respect to $\Delta$:
\be\label{deleq}\begin{split}
&0 = \Delta\bigg[\alpha_1 \cos^2\theta+\alpha_2\sin^2\theta
\\ &+\Delta^2\left(\beta_1\cos^4\theta
+\beta_2\sin^4\theta+2\beta_{12}\cos^2\theta\sin^2\theta\right)\bigg].
\end{split}\ee
We obtain
\be\label{min1}
\Delta_1^2 = -\alpha_1/\beta_1 \, , \quad \Delta_2^2 = 0
\ee
for $\theta = 0$ and
\be\label{min2}
\Delta_1^2 =0 \, , \quad \Delta_2^2 = -\alpha_2/\beta_2
\ee
for $\theta=\pi/2$.
We see that $\Delta/(2\pi T)$ is
indeed small near the second-order N-S$_i$ phase transition, since $\alpha_i\to 0$ at the transition.
  This also implies that in the superfluid state S$_i$ the chemical potential difference
is such that $|h_i|\ll 4\pi T$,
because the condition $|T- T_{c_i}|\ll T_{c_i}$, \eref{closeem},  makes the first term in the definition
\rref{alphaiem} of $\alpha_i$ small.

Having found the stationary points, we must check their stability.
For a minimum, the second derivative must be positive.  The second derivative
of the thermodynamic potential \rref{tpho} at stationary points  \rref{min1} and \rref{min2}
 vanishes when
\be\label{stabcond}
\alpha_i \beta_j -\alpha_j \beta_{12} =0 \, \qquad
(i\neq j).
\ee
These equations define the stability lines  enclosing the regions with two minima (gray areas in Fig.~\ref{fig:1}). Outside these
regions, there is only one minimum, while the other stationary point is a saddle.
Note that the fourth solution to \eref{theom} can be obtained by equating the terms in square brackets
to zero. This solution is present only when the thermodynamic potential has two minima
and corresponds to the saddle point between them.

Finally, the minima \re{min1} and \re{min2} are degenerate (the dashed lines in Fig.~\ref{fig:1})
when $\Omega(-\alpha_1/\beta_1,0)=\Omega(0,-\alpha_2/\beta_2)$, which yields
\be\label{degcon}
\frac{\alpha_1^2}{\beta_1} = \frac{\alpha_2^2}{\beta_2} \, .
\ee
This condition can be satisfied only close to both second-order N-S$_i$ transitions, so that
 $|h_i|\ll 4\pi T$ for both $i=1$ and $i=2$; see the discussion after \eref{min2}.
Substituting \esref{alphaiem} and \rref{betaiem}
into \eref{degcon} we obtain to leading order in $|h_i|/4\pi T$
\be\label{degapp}
\left[-\frac{1}{2}\Psi''\left(\frac12\right)\right] \left(\frac{h_1^2-h_2^2}{(4\pi T)^2}\right) =
 \ln \frac{T_{c_1}}{T_{c_2}} \, .
\ee
This equation shows that chemical potential differences, temperature,
and the asymmetry in the interaction strengths
determine the lines of the first-order phase transitions between different superfluid
states. At fixed $T$, \eref{degapp} defines transition lines $h_1(h_2)$ in the $h_1$-$h_2$ plane;
see the dashed curves in Fig.~\ref{fig:1}.

In the presence of a trapping potential $V(\rr)$, we can combine our phase diagram of Fig.~\ref{fig:1}
with the so-called LDA \cite{review} to predict the formation
of different superfluid shells in the trap. The LDA
 assumes position-dependent chemical potentials
\be\label{locapp}
\mu_i = \mu_i^0 - V(\rr)\, , \quad i=1,2,3 \, .
\ee
The differences $h_1=\mu_3^0-\mu_2^0$ and $h_2=\mu_3^0-\mu_1^0$ [see Eq.~\re{difem}] remain constant
throughout the trap and identify a point $\vec{h}=(h_1,h_2)$ on the
phase diagram (Fig.~\ref{fig:1}).  The temperature $T_{c_i}\equiv T_{c_i}(\mu_i)$ depends on the
chemical potential $\mu_i$ as in the standard BCS theory; see the text below Eq.~\re{b12em}.
As $\mu_i$ decreases from the center to the edge of the trap,  $T_{c_i}(\mu_i)$ also decreases.
On the other hand, the positions of the lines in the phase diagram in  Fig.~\ref{fig:1} are determined
by the values of  $T_{c_i}(\mu_i)$; see, e.g., \eref{degapp}. Therefore, the ``local''
phase diagram -- i.e., the phase diagram of the homogenous system that corresponds to
the values of chemical potentials at a particular point $\rr$ in the trap -- changes, and
as we move from its center towards the edge, the regions where
the superfluids are stable become smaller due to the decrease in $T_{c_i}(\mu_i)$. The
actual values of $\mu_i^0$ and consequently $h_i$ must be determined self-consistently by
fixing particle numbers for species 1,2, and 3. Depending on
the position of the resulting point $\vec{h}$ in the local phase diagram at the trap center,
different configurations are possible. For example, if $\vec{h}$ is in the S$_1$ region at the center,
the evolution of the local phase diagram with the position $\rr$ can bring this point into the N region
or make it pass through the S$_2$ region first. These two possibilities correspond to
a central superfluid  S$_1$ core surrounded by
a normal shell or a superfluid S$_1$ core followed by an S$_2$ shell and a normal shell farther out,
respectively. If $\vec{h}$ is in the S$_2$ region at the trap center, on the other hand, we obtain
a superfluid  S$_2$ core surrounded by a normal shell.
Alternatively, for low particle number
the normal-state atoms of the noncondensed species could form a normal core overlapping with
the superfluid one.
This qualitative picture is in agreement with numerical results of \cite{tor2}.
However, as we will discuss at the end of Sec.~\ref{sec:dw}, the LDA  has rather limited applicability
in the presence of an  S$_1$-S$_2$ boundary.

Let us summarize our observations so far in this section about the
possible phases and phase transitions in the homogeneous case. We
saw that for $T>T_{c_1}>T_{c_2}$ the system is in the normal state N
for any  chemical potentials differences  $h_1=\mu_3-\mu_2$ and
$h_2=\mu_3-\mu_1$ (recall that we set the coupling constant $g_3$
between species 1 and 2 to  zero, while $|g_1|>|g_2|$). A
second-order phase transition to the superfluid state S$_1$ where species
$2$ and $3$ condense first happens at $h_1=0$ and $T=T_{c_1}$ at arbitrary
$h_2$. For $T_{c_2}<T<T_{c_1}$ the only possible states are the
normal state and superfluid S$_1$. At lower temperatures $T<T_{c_2}$
three states can exist as shown in Fig.~\ref{fig:1}. A second-order transition
from the normal state to superfluid S$_2$ first takes place at
$T=T_{c_2}$, $h_2=0$, and sufficiently large $|h_1|$ (so that S$_2$
wins over S$_1$); see  Fig.~\ref{fig:1}. The   S$_1$-S$_2$ transition is
always first order, while  the N-S$_2$ and N-S$_1$ are both second
order provided that the temperature is above the tricritical
temperatures \re{triem}, i.e.,
\be\label{hight}
 T>\max\{
T^1_{\mathrm{tri}}, T^2_{\mathrm{tri}}\} \equiv T_{\mathrm{tri}}.
\ee
For $T< T_{\mathrm{tri}}$ at least one of the transitions N-S$_1$ or N-S$_2$
becomes first order and the Ginzburg-Landau
expansion \re{GLe} breaks down; see the discussion below
Eq.~\re{GLcond}.

\subsection{Phase separation}

In the previous subsections we analyzed the phase diagram in the grand-canonical ensemble.
Here we consider the canonical ensemble; i.e., we fix the densities $n_i$ of the three fermionic species.

The corresponding chemical potentials are found by solving the equations
\be
n_i = - \frac{\partial \Omega}{\partial \mu_i} \, .
\ee
If the differences between the densities are
large, the chemical potential differences are also large and
the gas is in the normal state. Let us assume that the densities deviate little from
an average density $n_0$,
\be
n_i = n_0 + \delta n_i \, .
\ee
Then, density deviations can be written as the sum of a noninteracting
term and a correction due to the presence of the superfluid:
\be\label{dneq}
\delta n_i = \nu \delta\mu_i - \frac{\partial \delta\Omega}{\partial \mu_i} \, ,
\ee
where $\delta\mu_i = \mu_i - \mu_0$, $\delta\Omega = \Omega-\Omega_N$, and
$\mu_0$ is the chemical potential for a noninteracting gas with density $n_0$.
In Eq.~\re{dneq} we neglected finite-temperature corrections to the
noninteracting contribution $\nu\delta\mu_i$ \cite{note1}.

\begin{figure}
\begin{flushleft}\includegraphics[width=0.45\textwidth]{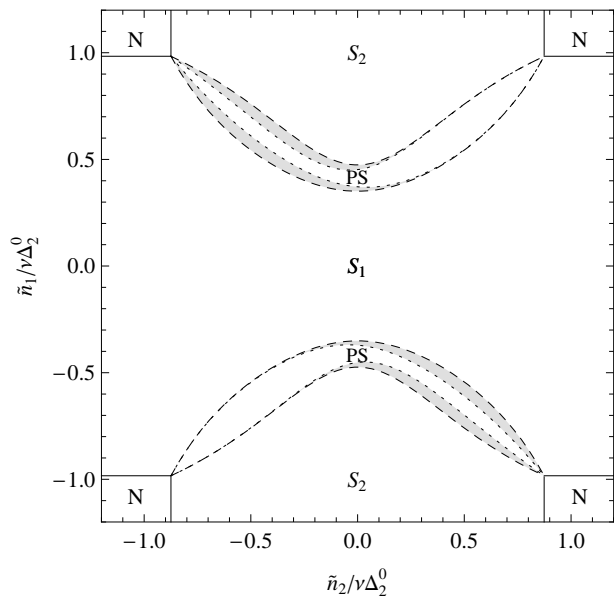}\end{flushleft}
\caption{Finite-temperature ($T=0.85T_{c_2}$) phase diagram for a three-component
Fermi gas in the plane of particle density
differences $\tilde{n}_1$-$\tilde{n}_2$, \eref{dnem}.
As in Fig.~\ref{fig:1}, the two nonvanishing pairwise couplings are chosen so that $T_{c_1}/T_{c_2}=1.04$.
The horizontal (vertical) segments denote the second-order phase transition  N-S$_1$ (N-S$_2$)
between normal (N) and superfluid state S$_1$ (S$_2$).
Dashed curves represent the limits of stability for the homogenous superfluids
and enclose
the phase-separated (PS) states. The shaded areas are the supercooling regions  where
a homogeneous superfluid state is metastable toward phase separation.}
\label{fig:1b}
\end{figure}

For example, if the system is in the superfluid state S$_1$, we find using \esref{dneq},
\rref{tpho}, and \rref{min1}:
\be\label{dnem}
\begin{split}
\tilde{n}_1\equiv n_3-n_2 &= \nu h_1- 2\frac{\partial\delta\Omega_1}{\partial h_1} \, ,
\\
\tilde{n}_2\equiv n_3-n_1 &= \nu h_2 \, ,
\end{split}
\ee
where
\be\label{omega1em}
\delta\Omega_1  = -\nu \frac{\alpha_1^2}{2\beta_1} \, .
\ee
Using similar equations for the homogenous superfluid S$_2$, we obtain the phase diagram
presented in Fig.~\ref{fig:1b} by mapping the lines in the phase diagram in the $h_1$-$h_2$
space of Fig.~\ref{fig:1} onto the corresponding lines in the $\tilde{n}_1$-$\tilde{n}_2$
space of density differences.
In particular, we note that each first-order phase transition line in the upper and lower half planes
of Fig.~\ref{fig:1} (dashed lines)
maps into two lines.  Indeed, according to Eqs.~\re{omega1em} and \re{dneq}, $\delta\Omega$ and
therefore $\tilde{n}_1$ and $\tilde{n}_2$ are different on the two sides of the transition.
This means that, as usual in the case of
first-order phase transitions, there is a region in the phase diagram where no homogeneous state
is stable and
phase separation must occur. Between this region and the stable homogeneous superfluid states, there are
supercooling regions (gray areas) where the homogeneous states are metastable towards phase
separation. The limits of these regions are found by mapping the corresponding
limiting metastability lines in the grand-canonical phase diagram
(dotted curves in Fig.~\ref{fig:1}).

At the end of the previous subsection we argued that, within the LDA, the two superfluid
states can coexist in a trap.
In this section we have shown that the transition
between the two superfluids is necessarily accompanied by a jump in the density. In this sense,
it is similar to the low-temperature transition between the superfluid and normal states in the
polarized two-component gas \cite{review}. In this system, the density jump signals a
potential breakdown
of the LDA on the length scale of the coherence length. There is also
evidence that surface tension effects should be taken into account to explain the shape of the
superfluid core in elongated traps \cite{muell}. In the present case of
the  S$_1$-S$_2$ transition   there are two competing
length scales (the two coherence lengths) that can affect the properties of the interface,
which we study in the next section.

\subsection{Phase diagram in the case when all three couplings are nonzero}
\label{g3em}

\begin{figure}
\begin{flushleft}\includegraphics[width=0.45\textwidth]{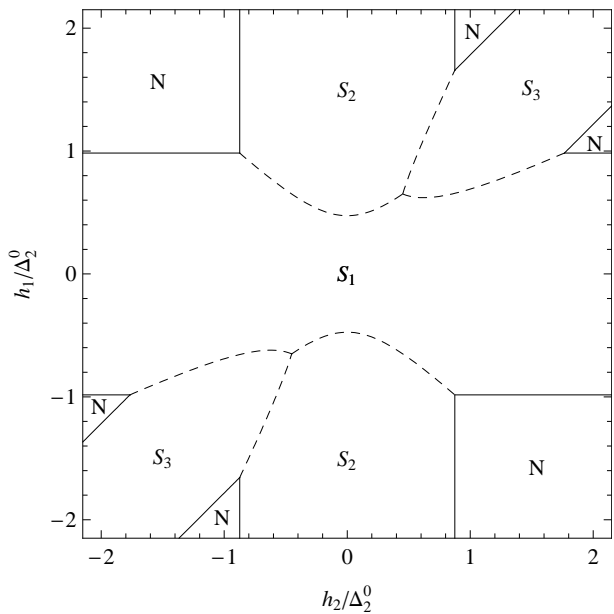}\end{flushleft}
\caption{Finite-temperature ($T=0.85T_{c_2}$) phase diagram for a three-component Fermi gas
with pairwise attraction between components in the plane of chemical potential
differences $h_1$-$h_2$, \eref{difem}. All three pairwise couplings are finite and such
that $T_{c_1}/T_{c_2}=1.04$ and $T_{c_3}/T_{c_2}=0.97$. As in the case of only two nonvanishing couplings
(cf. Fig.~\ref{fig:1}),
we identify the regions where the normal state (N) and the superfluid states (S$_{i}$) are
stable. The solid segments denote second-order N-S$_i$ transitions. The
dashed curves mark first-order S$_i$-S$_j$ transitions. Note that at the two points where these curves meet
all three superfluids can coexist.}
\label{fig:g3}
\end{figure}

Here  we briefly discuss the case when the  coupling constant $g_3$ between species 1 and 2 is also nonzero. Let
$|g_3|<|g_2|$. Now, in addition to $T_{c_{1,2}}$ there is the third temperature scale $T_{c_3}$. Similarly
to $T_{c_{1,2}}$,  it is defined as the critical
temperature of the superfluid with components 1 and 2 in the absence of 3.
Further, additional terms containing $|\Delta_3|^2$  appear in the thermodynamic potential
\rref{tpho}. The coefficient $\alpha_3$ of $|\Delta_3|^2$ is
defined by \eref{alphaiem} with $h_3=\mu_2-\mu_1 =h_2-h_1$.
For $T>T_{c_3}$, we have $\alpha_3>0$ and the phase diagram in Fig.~\ref{fig:1} is unchanged.
For $T_\mathrm{tri}<T<T_{c_3}$, new N-S$_3$ second-order phase transitions are possible as
well as first-order transitions S$_3$-S$_1$ and S$_3$-S$_2$. A phase diagram
with these transitions is shown in Fig.~\ref{fig:g3}. Note that, since
$h_3  =h_2-h_1$ is not an independent parameter, the phase diagram for the general case $g_3\ne 0$ can be plotted in the same $h_1$-$h_2$ plane as before.

\section{Domain wall}
\label{sec:dw}

Until now we considered a spatially uniform system where one of the
 phases occupies the entire space.
On the other hand, we have seen in the previous section that for a certain range of densities
phase separation of the two superfluids S$_1$ and S$_2$ occurs, as shown in Fig.~\ref{fig:1b}.
This implies the formation of  domain walls between homogeneous phases.  Similarly, domain walls must form at the
boundaries between S$_1$ and S$_2$  in a trapped three-component gas; see the text below Eq.~\re{locapp}.
Let us analyze the properties of the domain wall using a grand-canonical thermodynamic potential.
Its minima   that correspond to the homogenous states S$_1$ and S$_2$
far from the domain wall must be degenerate for the superfluids to coexist in between. Indeed, the entire
phase-separated
regions in the diagram in Fig.~\ref{fig:1b} correspond to the lines of degenerate minima in Fig.~\ref{fig:1} (dashed curves).
To obtain the domain wall solution, we need to retain the
gradient terms in the
thermodynamic potential, \eref{GLe}, and minimize it subject to appropriate boundary condition.
We first consider temperatures close to the critical
one and later  extend our considerations to lower temperatures.

\subsection{Domain walls at temperatures close to the critical ones}
\label{sec:dwtc}

Here, as in Sec.~\ref{sec:pdtc}, we assume that
conditions \re{closeem} hold.
As discussed below \eref{degcon}, in this case the chemical
potential differences are such that
$|h_i|/4\pi T\ll 1$.
Then, the prefactors in front of the gradient terms in \eref{GLe} can both be approximated with
\be\label{xi0}
\xi_0^2 = \frac{7\zeta(3)}{12}\left(\frac{v_F}{2\pi T}\right)^2 \, .
\ee
We note that we cannot
neglect the small differences of order $(h_i/2\pi T)^2$ in the prefactors of the fourth-order terms
$\beta_i$ and $\beta_{12}$, as these differences enter into the equations that determine
the value of the order parameter; see \eref{theom}.

For simplicity, let us assume that the translational invariance is broken only along the $x$ axis, so that
the system is in the homogeneous state S$_1$ at $x\to -\infty$ and in the state S$_2$ at $x\to +\infty$.
According to Eqs.~\re{min1} and \re{min2}, this means $\theta\to 0$ for $x\to -\infty$, $\theta\to \pi/2$ for $x\to +\infty$,
and $d\theta/dx \to 0$ for $x\to \pm\infty$.
The minimization of the thermodynamic potential \re{GLe} yields a system of two
second-order nonlinear differential equation for $\Delta(x)$ and $\theta(x)$ defined in \eref{poldec}.
These equations admit a first integral, the
conserved ``energy'' of the domain wall:
\be\label{dwen}\begin{split}
&
-\xi_0^2 \left(\nabla \Delta\right)^2 -
\xi_0^2 \Delta^2 \left(\nabla \theta \right)^2
+ \Delta^2 \Big(\alpha_1 \cos^2\theta +\alpha_2 \sin^2\theta\Big)
\\ &
+ \frac{\Delta^4}{2}\Big(\beta_1 \cos^4\theta + \beta_2 \sin^4 \theta +
2 \beta_{12}\cos^2\theta\sin^2\theta\Big)\, .
\end{split}\ee
Our assumption \re{closeem} implies that the two homogenous states S$_1$ and S$_2$  have close
values of the order parameter amplitude $\Delta$; see \esref{min1} and \rref{min2}. This enables us  to neglect
the $(\nabla\Delta)^2$ term in \eref{dwen}.
This term changes little on the length scale associated with the width of the
domain wall, while the angular
variable $\theta(x)$ changes by $\pi/2$ on the same length scale; i.e., the ratio of the $(\nabla\Delta)^2$ and
$(\nabla\theta)^2$   terms in \eref{dwen}
is of order $(T_{c_1}-T_{c_2})/T_{c_1}$. Solving
\eref{deleq} for $\Delta$ in terms of $\theta$ and substituting the result into
\eref{dwen}, we arrive at
\be\label{dweq1}\begin{split}
& \left[\left(\xi_0\nabla\theta\right)^2 -\frac{1}{2}
\Big(\alpha_1 \cos^2\theta +\alpha_2 \sin^2\theta\Big) \right]
\\ & \times \frac{\alpha_1 \cos^2\theta +\alpha_2 \sin^2\theta}
{\beta_1 \cos^4\theta + \beta_2 \sin^4 \theta +
2 \beta_{12}\cos^2\theta\sin^2\theta} = -S \, .
\end{split}\ee
The value of the constant $S$
on the right-hand side can be determined from the boundary conditions
$\theta \to 0$ and $d\theta/dx \to 0$ as $x \to -\infty$:
\be\label{sval}
S = \frac{\alpha_1^2}{2\beta_1} \, .
\ee
Note that $\nu S$ is the condensation energy density for the homogenous state.
Indeed substituting, e.g., \eref{min1} into \eref{tpho} we obtain $\Omega_N -\Omega = \nu S$;
see also \eref{omega1em}.

Using \eref{degcon}, we rewrite \eref{dweq1} as
\be\label{dweq2}
\frac{2 d\theta}{dx} = \frac{1}{\ell} \frac{\sin 2\theta}{\sqrt{1+a\cos 2\theta}}\, ,
\ee
where
\be\label{adef}
a=\frac{\alpha_1-\alpha_2}{\alpha_1+\alpha_2}
\ee
 and
\be\label{ldef}
\ell^2 = \frac{1}{2}\left(\xi_1^2+\xi_2^2\right) \eta^2 \, .
\ee
Here we have introduced the coherence lengths of the two condensates,
\be\label{xitc}
\xi_i = \xi_0/\sqrt{-\alpha_i} \, , \quad i=1,2
\ee
and the scale factor
\be\label{etadef}
\eta^{-2} =\frac{\beta_{12}}{\sqrt{\beta_1\beta_2}}-1 \, ,
\ee
where $\alpha_i$, $\beta_i$, and $\beta_{12}$ are defined by Eqs.~(\ref{alphaiem}),
(\ref{betaiem}), and \re{b12em}, respectively.
In particular, to leading order in $h_i/4\pi T$ we have
\be\label{etatc}
\eta \simeq \sqrt{12 \frac{\Psi''\left(\frac12\right)}{\Psi^{(4)}\left(\frac12\right)}}
\frac{4\pi T}{h_1 - h_2}
\simeq 0.512 \frac{4\pi T}{h_1 - h_2} \, .
\ee

From \eref{dweq2}, we obtain an implicit equation for the spatial
dependence of $\theta$:
\be\label{shapeem}\begin{split}
\frac{x-x_0}{\ell} = & \sqrt{1-a}\, \mathrm{arctanh}\left[\sqrt{\frac{1-a}{1+a\cos 2\theta}}\right]
\\ & - \sqrt{1+a}\, \mathrm{arctanh}\left[\sqrt{\frac{1+a\cos 2\theta}{1+a}}\right].
\end{split}\ee
The parameters $a$ and $\ell$ characterize the asymmetry of the domain wall
with respect to reflection $(x-x_0) \to - (x-x_0)$ and
its width, respectively. The parameter $\ell$ provides a new length scale, in addition to the coherence lengths,
via the (large) parameter $\eta$; see \eref{ldef}. In the next subsection, we will see
that the same parameter also enters the expression for the surface tension associated
with the domain wall.

\begin{figure}
\begin{flushleft}\includegraphics[width=0.47\textwidth]{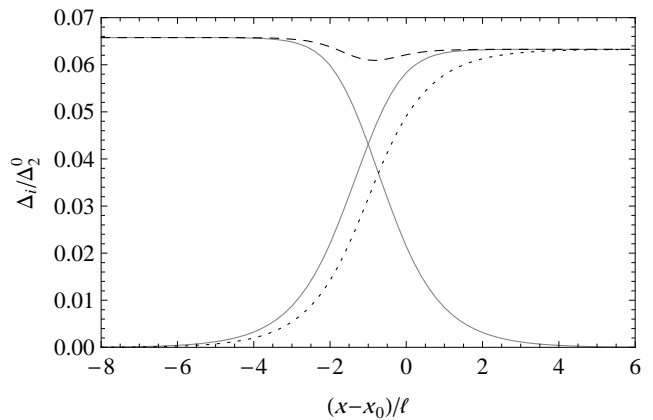}\end{flushleft}
\caption{Profiles of the order parameter components $\Delta_1$ (decreasing solid line)
and $\Delta_2$ (increasing) in the presence of a domain wall between two superfluid states
of a three-component Fermi gas. Here
$T=0.92T_{c_2}$, $T_{c_1}/T_{c_2}=1.05$, and the chemical potential differences are
$h_2/\Delta_2^0=-0.66$ and
$h_1/\Delta_2^0\simeq 0.84$, where $T_{c_i}$ and $\Delta_i^0$ are defined below \eref{b12em}.
Note the overlap of the two components over the central region of size $\ell$, \eref{ldef}.
We also show the order parameter in the polar decomposition of \eref{poldec};
the dashed line is used for $\Delta$ and the dotted line for $\theta\times[0.063/(\pi/2)]$.}
\label{fig:2}
\end{figure}

An example of the spatially nonuniform order parameters $\Delta_1(x)$ and
$\Delta_2(x)$ in the presence of a domain wall is shown in Fig.~\ref{fig:2}.
We also plot the angular variable $\theta(x)$ (rescaled) and
the amplitude $\Delta(x)$. Note that $\Delta(x)$ shows little change. This is consistent with the assumption
that gradients of $\Delta(x)$ can be neglected near the critical temperature $T_{c_1}$.

Finally, we note that because the densities on the two sides of
the domain wall are different [see Eq.~\re{dnem}], it could, in principle, be detected by imaging the sample.
For bosonic atoms, overlap between two Bose-Einstein condensates was observed long ago
\cite{cornell} (for theoretical studies of the two-component bosonic system, see Ref.~[\onlinecite{barankov}]).
Alternatively, spatially resolved rf spectroscopy \cite{srrf} could reveal
the different gaps.

\subsection{Surface tension}

From the point of view of thermodynamic properties, the presence of a surface
separating the two condensates can be taken into account by including a surface tension term
in the thermodynamic potential \cite{LL5}. Moreover, as  mentioned
above, a proper treatment of surface tension effects is necessary to  describe correctly the
condensate profile in asymmetric traps \cite{muell}.

The surface tension
$\sigma$ can be calculated by integrating the
difference between the potential in the presence of the domain wall ($\Omega_{\mathrm{dw}}$) and
the one in the uniform state ($\Omega_{\mathrm{u}}$) over
the direction perpendicular to the domain wall:
\be
\sigma = \int\!dx \, \Big(\Omega_{\mathrm{dw}} - \Omega_{\mathrm{u}}\Big) \, .
\ee

Using \eref{dweq2}, we derive
\be\label{ste1}
\begin{split}
\sigma & = \frac{2\nu S \ell}{\eta^{2}} \int\!d\theta
\frac{4\sqrt{\beta_1\beta_2}\sqrt{1+a \cos 2\theta}\sin 2\theta}{b_- \cos^2 2\theta +2b_{12}
\cos 2\theta + b_+}
\\ & \equiv \frac{2\nu S \ell}{\eta^{2}} f(a,\{\beta\}) \, ,
\end{split}
\ee
with $S$, $\ell$, and $\eta$ defined in \esref{sval}, \rref{ldef}, and \rref{etadef}, respectively,
\be
b_\pm = \beta_1+\beta_2 \pm 2\beta_{12} \, , \quad b_{12} = \beta_1 - \beta_2 \, ,
\ee
and $\beta_1$, $\beta_2$, and $\beta_{12}$ defined in \esref{betaiem} and \rref{b12em}. Using these
definitions and  $h_i/4\pi T\ll 1$ [see the text below   \eref{degcon}], we estimate
$b_+\sim O(1)$, $\beta_i \sim O(1)$, $b_{-} \sim O(h^2/T^2)$, and $b_{12} \sim O(h^2/T^2)$.
Therefore, we replace
the denominator in the integral \re{ste1} by $b_+$ and obtain
\be
f(a,\{\beta\}) \simeq \frac{1}{3a} \left[\left(\sqrt{1+a}\right)^3-\left(\sqrt{1-a}\right)^3\right] \, .
\ee
By definition \re{adef}, the asymmetry parameter varies between $-1$ and $1$. Therefore,
$1 \ge f \ge 2^{3/2}/3 \simeq 0.94$.
Neglecting this weak dependence on the asymmetry $a$, we can write
\be\label{stapp}
\sigma \simeq \nu S \sqrt{2\left(\xi_1^2+\xi_2^2\right)}\,\eta^{-1} \, .
\ee
This expression shows that the surface tension is determined by the value of the condensation energy
$\nu S$ for the
uniform system times the (root-mean-square) coherence length divided by the scale factor.
As we will see in the next subsection, this formula for the surface tension is valid
in a wider range of temperatures than the limiting case $T_{c_2}-T \ll T_{c_2}$ considered here.

\subsection{Intermediate temperatures}
\label{sec:var}

In the preceding subsections we have considered a domain wall near the critical temperature.
On the other hand, as discussed at the end of Sec.~\ref{sec:pdtc}, the Ginzburg-Landau approach remains
generally valid near
second-order phase transitions even at lower temperatures above $T_\mathrm{tri}$; see \eref{hight}.
So we can in principle analyze the properties of the domain wall at intermediate temperatures
(and for larger differences in the critical temperatures than in the previous subsections).
Approaching $T_\mathrm{tri}$, the parameter $\beta_1$ becomes small by definition, while
in the superfluid state $\alpha_1$ is finite, so we expect the difference between $\Delta_1$ and
$\Delta_2$ to grow; see \esref{min1} and \rref{min2}.
If this is the case, the approximation in which the
gradient of $\Delta$ is neglected breaks down.
To remedy this, we construct in this
section a variational domain wall solution.

As a starting point for the variational approach, we note that the approximate domain wall
solution is determined by three parameters: the position $x_0$, the asymmetry $a$, and the
size $\ell$. The first one cannot affect the energy (surface tension), as it only reflects the
translational invariance of the infinite system, and henceforth we  set $x_0=0$.
In the (unphysical \cite{noteu}) symmetric limit $a\to 0$, we can
obtain an explicit expression for, e.g., the profile of $\Delta_1$:
\be
\Delta_1 = \Delta \sqrt{\frac{1}{2}\left[1-\tanh\left(\frac{x}{\ell}\right)\right]} \, .
\ee
This suggests the following trial functions for the order parameters:
\be
\Delta_i = \sqrt{\frac{-\alpha_i}{\beta_i}}
\sqrt{\frac{1}{2}\left[1\mp\tanh\left(\frac{x}{\ell_\mathrm{v}}\mp \frac{\delta}{2}\right)\right]} \, ,
\ee
where the $\Delta_i$ are fixed to their asymptotic values at $x\to\pm\infty$. We  introduced a parameter
$\delta$ which describes the overlap between the two superfluids and enables us to take into account
the role of the interaction term in \eref{GLe}.
As before, we also have a parameter $\ell_\mathrm{v}$ related to the domain wall thickness \cite{notel}.
Both parameters
must be determined by minimizing the surface tension:
\be\begin{split}
\sigma = & \nu S \ell_\mathrm{v}\left[-1-\delta+\frac{\beta_{12}}{\sqrt{\beta_1\beta_2}}
\delta\Big(1+\coth(\delta)\Big)\right] \\
& + \frac{\nu S}{2\ell_\mathrm{v}}\left(\tilde{\xi}_1^2+\tilde{\xi}_2^2\right) \,
\end{split}\ee
with the coherence lengths
\be\label{xitil}
\tilde{\xi}_i^2 = \frac{v_F^2}{3}\frac{\beta_i}{-\alpha_i} \, ,
\ee
which reduce to \eref{xitc} as $T \to T_{c_i}$.

After minimization, $\sigma$ can be written as
\be\label{stvar}
\sigma = \nu S \sqrt{2\left(\tilde{\xi}_1^2+\tilde{\xi}_2^2\right)}\,\eta_{\mathrm{v}}^{-1},
\ee
with the variational scale parameter given by
\be
\eta_\mathrm{v}^{-2} = \delta_0\Big(1+\coth(\delta_0)\Big)\frac{\beta_{12}}{\sqrt{\beta_1\beta_2}}
-\left(1+\delta_0\right) \, ,
\ee
where $\delta_0$ is the solution to
\be
\coth\delta_0 - \frac{\delta_0}{\sinh^2\delta_0}+1 = \frac{\sqrt{\beta_1\beta_2}}{\beta_{12}} \, .
\ee
Note that near the critical temperature, the right hand side of the above equation tends to unity,
so that $\delta_0 \to 0$ and $\eta_\mathrm{v} \to \eta$. Since the variational approach gives an
upper bound on the surface tension, it also gives a lower one on the domain wall thickness:
\be\label{lvdef}
\ell_\mathrm{v} = \sqrt{\frac{1}{2}\left(\tilde{\xi}_1^2+\tilde{\xi}_2^2\right)}\, \eta_\mathrm{v} \, .
\ee

\begin{figure}
\begin{flushleft}\includegraphics[width=0.47\textwidth]{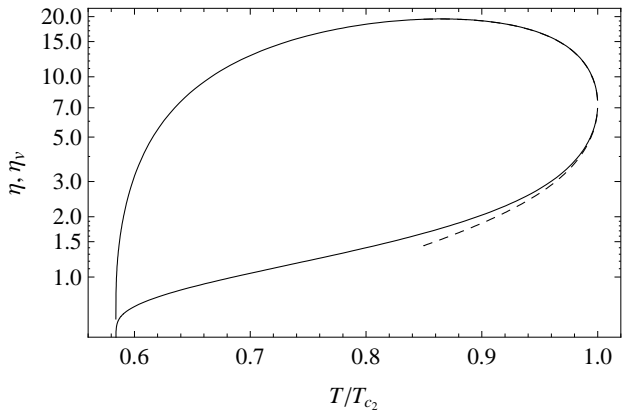}\end{flushleft}
\caption{Temperature dependence of the scale factors $\eta_\mathrm{v}$ (solid lines) and $\eta$ (dashed lines)
relating the domain wall size and the coherence lengths; see \esref{lvdef} and \rref{ldef}. The coupling
constants are chosen so that $T_{c_1}/T_{c_2}=1.04$.
The chemical potential differences $h_i$ are the critical ones; see the text after \eref{lvdef}.
They have opposite signs for the lower curves [$h_1=\pm h_1^c(T)$, $h_2 =\mp h_2^c(T)$] and
the same sign for the upper ones [$h_1=\pm h_1^c(T)$, $h_2 =\pm h_2^c(T)$]. In the latter case
the difference between the two scale factors is not visible. Note that the scale factors are
larger near $T_{c_2}$ and/or for chemical potential differences of the same sign.}
\label{fig:3}
\end{figure}

In Fig.~\ref{fig:3} we compare the behavior of the scale factors $\eta$ and $\eta_\mathrm{v}$ as functions
of temperature for $T_{c_1}/T_{c_2}=1.04$.  The curves are computed in the two limiting cases in which
the chemical potential differences $h_i$ are the critical ones, $h_i = \pm h_i^c(T)$;
cf. the discussion after \eref{nsstab}. They correspond to the points where the first-order
transition lines meet the second-order ones in Fig.~\ref{fig:1}. There are two inequivalent cases
depending on the relative sign between $h_1$ and $h_2$. We choose these points
in the phase diagrams because the Ginzburg-Landau expansion \rref{GLe} is always valid in their vicinity
as long as the temperature is above the tricritical temperature \re{hight}; see Sec.~\ref{glexp}. At fixed temperatures close to $T_{c_2}$, we find
that the scale parameter $\eta$ evolves smoothly as a function of the chemical
potential differences along the first-order transition lines -- i.e.,
going from one limiting case to the other one.
Therefore, the limiting cases displayed in Fig.~\ref{fig:3} give upper and lower bounds on the possible values of
the variational scale parameter $\eta_\mathrm{v}$.

Note that close to $T_{c_2}$ the two parameters $\eta_\mathrm{v}$ and $\eta$
have very similar values, and the approximate solution can be trusted in this regime. At smaller
temperatures and chemical potential differences of opposite signs,
the approximations made in the previous sections become invalid, and
$\eta$ decreases more rapidly than $\eta_\mathrm{v}$. The latter remains of order
unity at intermediate temperatures before quickly decreasing near $T_\mathrm{tri}^1$.
On the contrary, for chemical potential differences of the same sign, both approaches give
similar results. Moreover, $\eta_\mathrm{v}$ initially {\it increases} with decreasing temperature,
leading to potentially very thick domain walls with significant overlap between the two
superfluid states. Again, when approaching $T_\mathrm{tri}^1$ the scale parameter
$\eta_\mathrm{v}$ quickly decreases. However, at these temperatures
the present approach is invalid -- higher orders in the Ginzburg-Landau expansion become relevant.

The above observations show the limits of applicability of the LDA
\re{locapp} in the presence of a trap. For large $\eta_\mathrm{v}$, the surface tension
is small, and we expect the density profiles to follow the shape of the trapping potential.
On the other hand, in this case the width $\ell_\mathrm{v}$ of the domain wall is large; see \eref{lvdef}.
The order parameter components vary smoothly on this scale and
density jumps predicted by the LDA cannot be a good approximation of the actual density profiles.
In other words, the LDA breaks down, not on a length scale $\tilde{\xi}_i$, as usually assumed,
but on a much longer scale.
In the opposite case of small $\eta_\mathrm{v}$,
the situation is reversed: the densities vary quickly on a length scale comparable to
the coherence lengths. However, now the surface tension becomes important in asymmetric traps and the
densities do not simply follow the profile of the trapping potential as in the LDA. This is seen, e.g.,
in the polarized two-component gas at low temperatures \cite{muell}.

The above considerations are valid under the assumption than the sample size $R$ is much larger
than the domain wall thickness $\ell$ \cite{note1b}, in which case finite-size effects can be neglected. This requirement also
limits the validity of the LDA.
We can estimate how large, in terms of the number $N$ of trapped atoms for each species, the sample
should be in order to
accommodate a domain wall. In the weak-coupling limit, a good estimate of the sample size is given
by the Thomas-Fermi radius
\be\label{tfem}
R \simeq a_{ho} (48 N)^{1/6} \, ,
\ee
where $a_{ho} = \sqrt{1/m\omega_{ho}}$ is the harmonic oscillator length in the parabolic trap
$V(r) = \frac{1}{2}m\omega_{ho}^2r^2$. Next, we estimate the value of the coefficient $\xi_0$,
\eref{xi0}, at $T\simeq T_{c_2}$ using
\be\label{tctr}
T_{c_2} \simeq 0.28 E_F e^{-\pi/2 k_F |a_s|} \, ,
\ee
where $a_s$ is the (negative) scattering length and the Fermi momentum (at the trap center) is
\be\label{kfem}
k_F \simeq \frac{1}{a_{ho}}(48N)^{1/6} \, .
\ee
The previous three expressions (\ref{tfem}), (\ref{tctr}), and \re{kfem} can be found in
Ref.~[\onlinecite{review}]. Substituting \eref{tctr} into
\eref{xi0}, we find
\be\label{xi0as}
\xi_0 \simeq \frac{1}{k_F} e^{\pi/2k_F |a_s|} \, ,
\ee
which is a lower bound for the coherence lengths defined in \eref{xitc}.
Then, for $\ell$, \eref{ldef}, we can write
\be
\ell \gtrsim \frac{1}{k_F}e^{\pi/2k_F |a_s|} \, \eta \, .
\ee
For $k_F |a_s| \simeq 1$, the requirement $R/\ell \gg 1$ in terms of the
total particle number $N_t \sim 3N$ becomes
\be\label{Nest}
N_t^{1/3} \gg 2\eta \, .
\ee
For a scale factor $\eta \sim 10$, this gives $N_t\gg 10^4$.
However, due to the slow growth with $N_t$ of
the left-hand side of \eref{Nest}, even for a typical sample size
with $N_t\sim 10^7$ \cite{srrf} the ratio $R/\ell \sim 10$, and finite-size
effects should be taken into account.
Note that these estimates are also sensitive to
the interaction strength $k_f |a_s|$ [cf. \eref{xi0as}], and for a weaker interaction
(e.g., $k_F |a_s| \simeq 0.5$) we obtain $N_t^{1/3} \gg 9 \eta$ instead of \eref{Nest} and $N_t\gg 10^6$.

\section{Phase diagram at $T=0$}
\label{ztpd}

All our previous consideration have been restricted to the vicinity
of second-order phase transitions and hence to the
``high-temperature'' regime $T>T^1_{\mathrm{tri}}$. There is also a simple
explicit description of the phase diagram at zero temperature, which
we present in this section. In this case, all phase transitions
(N-S$_{i}$ and S$_{1}$-S$_{2}$) are first order, and the components
of the order parameter $\vec{\Delta}$ at the minima of the
thermodynamic potential are either zero or independent of the
chemical potential differences \cite{colsup}. Our phase diagram is
in qualitative agreement with the numerical results of
Refs.~[\onlinecite{tor1,crossover1}]. Although we
consider the weak-coupling regime, we expect that our results will
not qualitatively change at stronger coupling on the BCS side of the
crossover. On the BEC side, on the other hand, the system behaves as
a Bose-Fermi mixture (see, e.g., \cite{ps} for the two-component
system), and we cannot exclude the possibility of qualitative
differences  (see also \cite{crossover2}). Finally, we note that in
constructing the zero-temperature phase diagram we consider for
simplicity only uniform states, neglecting the possibility that a
spatially varying order parameter may be energetically favored in
some regions of the phase diagram, as is the case for the FFLO
state \cite{FFLO} in a two-component system \cite{FFLOnote}; see,
e.g., Refs.~\cite{colsup,rad,FFLO1,FFLO2,FFLO3}.

\begin{figure}
\includegraphics[width=0.44\textwidth]{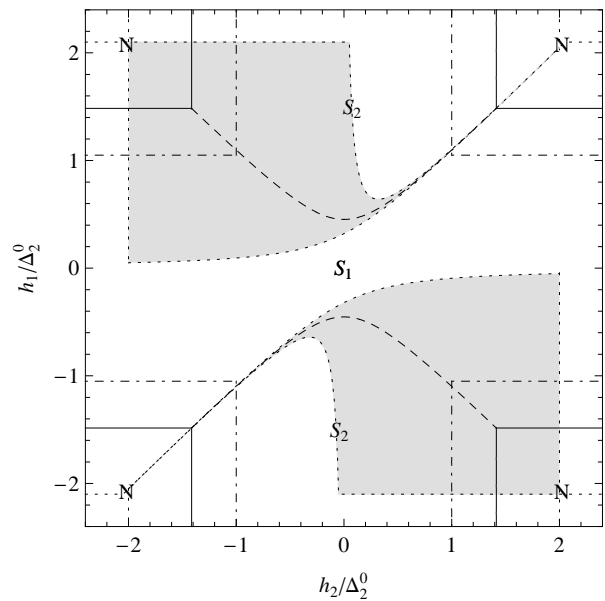}
\caption{Zero-temperature phase diagram for a three-component Fermi gas in the plane $h_1$-$h_2$ of chemical
potential differences, \eref{difem}. The two nonvanishing couplings constants are
such that   $\Delta_0^1/\Delta_0^2=1.05$, where $\Delta_0^i$ are defined below \eref{tpeq}.
As at high temperature (see Fig.~\ref{fig:1}), the normal state (N) is stable for large
$h_i$. Horizontal (vertical) solid segments denote first-order N-S$_1$ (N-S$_2$)
transitions between normal and superfluid states
(in contrast, at high temperatures these
transitions are second order). The dashed curves identify first-order S$_1$-S$_2$ transitions.
The dotted curves represent the superfluid-state stability limits and the dot-dashed lines the
normal-state stability limits. Shaded areas are regions where both superfluid states are (meta)stable.
Note that these regions are much larger than the corresponding ones at high temperature in Fig.~\ref{fig:1} and overlap
with the normal-state stability regions.}
\label{fig:4}
\end{figure}

According to \eref{tpeq}, the differences $\delta\Omega_i$ between the thermodynamic
potentials in the condensed and
the normal states at the same chemical potentials  is
\be\label{zttp}
\delta\Omega_i=-\frac{1}{2} |\Delta_i^0|^2 + \left(\frac{h_i}{2}\right)^2,\quad i=1,2 \, .
\ee
Equating $\delta\Omega_i$ to zero, we obtain the first-order transition lines
between  superfluids S$_i$ and the normal state (the Clogston-Chandrasekhar \cite{CC} critical field). The condition
 $\delta\Omega_1=\delta\Omega_2$ yields
the first-order transition line between the two condensates. These transitions are plotted as solid
and dashed lines, respectively, in Fig.~\ref{fig:4}.

Considering as before the quadratic fluctuations [cf. \eref{stabcond}; see also the next section], we
determine the zero-temperature instability lines
\be\label{ztsc}
h_j \left(h_i-h_j\right)
=|\Delta_i^0|^2-|\Delta_j^0|^2 \, , \quad i,j = 1,2,
\ee
for S$_i$ becoming unstable towards S$_j$. Using these expressions,
we obtain the (dotted) stability curves in the phase diagram shown in Fig.~\ref{fig:4}.
The horizontal and vertical dotted lines indicate the instabilities of the
superfluid states towards the normal state, which are identified by the conditions \cite{colsup}
\be
h_i = 2\Delta_i^0 \, ,
\ee
while the dash-dotted lines mark the instability of the normal state, obtained from the
$T\to 0$ limit of \eref{nsstab}:
\be
h_i = \Delta_i^0 \, .
\ee

The resulting zero-temperature phase diagram shown in Fig.~\ref{fig:4} has a richer structure than that at ``high''
temperatures; see Fig.~\ref{fig:1}. For example, a larger region of the phase diagram
is occupied by metastable states due to the first-order nature of the transition
to the normal state. This in turn means that in the $\tilde{n}_1$-$\tilde{n}_2$ density-space
phase-separated states occupy a larger region
of the phase diagram. Consequently, more complicated domain
wall structures are possible that interpolate between the different superfluid states and the normal state as
well, as is the case for a two-component Fermi gas \cite{muell}. As remarked before,
the phase separation translates into density jumps in the LDA treatment of
the trapping potential. However, the validity of the LDA should be
confirmed by estimating the effects of domain walls and surface tension, as in the
finite-temperature case.

\section{Collective modes}
\label{sec:cm}

In the absence of an external potential, the existence of thick domain walls is a manifestation
of the presence of soft collective
modes. While the former are possible only in the presence
of degenerate ground states, the
latter are a more general feature of the multicomponent Fermi gas. In this section we present
the dispersion relations for these modes and comment on their role in limiting the
applicability of the BCS mean-field approach. In trapped Fermi gases the collective
modes are known to affect the experimentally accessible (hydrodynamiclike) response of the
system \cite{bp}.

For concreteness we assume that the ground state is the superfluid $S_1$ with homogeneous
order parameter $\Delta_1$ and consider small fluctuations around this state:
\be\label{del1fl}
\Delta_1 (\rr,t) = \Delta_1 (1+\psi(\rr,t))e^{\mathrm{i}\phi(\rr,t)} + \delta\Delta_2(\rr,t)\, .
\ee
The phase fluctuations described by $\phi$ correspond to the well-known
soundlike Anderson-Bogoliubov mode \cite{ABmode}, while the amplitude fluctuations $\psi$
have a mass equal to $2\Delta_1$ \cite{VL}. These two modes
have also been studied in the BCS-BEC crossover \cite{ERS}.
Here we are interested in the
fluctuations $\delta\Delta_2(\rr,t)$ due to pairing in the noncondensed channel.

The propagator $D(\omega,q)$ of the $\delta\Delta_2(\rr,t)$ field
is obtained by expanding \eref{effac} around the stationary point with $\Delta_1 \ne 0$, $\Delta_2 =0$:
\be\label{prop}\begin{split}
& \left[\nu D(\omega,q)\right]^{-1} =
 \ln\frac{\Delta_1^0}{\Delta_2^0} \\
 & +\left(\omega + h_1-h_2\right)
H(\omega, q; h_1,h_2) + J(\omega,q; h_1,h_2) \, ,
\end{split}\ee
where functions $H$ and $J$ are given in the Appendix and $\Delta_i^0$ are the
values of the zero-temperature order parameter components; see the text below \eref{effac}.
Here we concentrate on the cases of zero temperature and vicinity to
second-order phase transitions. Moreover, we consider only long wavelength
fluctuations -- i.e., $q \to 0$.

\subsection{Collective modes at $T=0$}

In the limit $T\to 0$, the propagator in \eref{prop} becomes
\be\label{prop0T}\begin{split}
& \left[\nu D_0\right]^{-1}=  \ln\frac{\Delta_1^0}{\Delta_2^0}+ {\cal H}(\omega) +
\\ & \left[\left(h_1-h_2\right)\frac{1}{2}\frac{\partial^2}{\partial h_2^2}-
\frac{\partial}{\partial h_2}\right]\left(\frac{1}{h_1-h_2}{\cal H}(0)\right)\frac{v_F^2 q^2}{3}\, ,
\end{split}\ee
with
\be
{\cal H}(\omega) = \frac{1}{2}\ln\left[1+\left(\frac{h_2-h_1/2}{\Delta_1^0}\right)^2-
\left(\frac{\omega+h_1/2}{\Delta_1^0}\right)^2\right].
\ee
Note that
\be\begin{split}
\left[ \nu D_0(0,0)\right]^{-1}\equiv & \frac12 \ln \left[\left(\Delta_1^0\right)^2+h_2^2-h_2 h_1\right]
\\ & \qquad \quad -\ln \Delta_2^0 =0
\end{split}\ee
yields the stability condition \re{ztsc} for $i=1$.
Indeed, since the contribution of $\delta\Delta_2$ fluctuations to the action is
$\delta\Delta_2^*(\omega,q) \left[D(\omega,q)\right]^{-1} \delta\Delta_2(\omega,q)$, the
condition $\left[D(0,0)\right]^{-1}>0$ determines the stability of the superfluid S$_1$ with respect to static uniform
fluctuations.

Consider, e.g., the stability (or lack of it) of the superfluid S$_1$ with respect
to shifts in the chemical potentials in the case of equal interaction strengths.
For $h_1=0$ and small $h_2$ we get
\be
\left[\nu D(0,0)\right]^{-1} \simeq \frac{1}{2} \left(\frac{h_2}{\Delta_1^0}\right)^2 > 0 \, ,
\ee
which shows that the superfluid S$_1$ is stable, as expected, since fluctuations toward condensation in the
1-3 channel need to overcome the ``Zeeman energy''; cf. \eref{zttp}. This is contrary to the claim
in Ref.~[\onlinecite{ssb}] that this chemical potential shift causes the system to become unstable.
In contrast, for $h_2=0$ the inverse propagator $D(0,0)^{-1}$ is zero for any $h_1$, which
indicates an instability. In this case the stable state is the superfluid S$_2$, as
can be seen by repeating the above analysis with $1 \leftrightarrow 2$.

Now let us determine the dispersion relation of collective modes. For simplicity, we
consider the case $h_1=0$. We have
\be
\omega^2 = m^2 + v_0^2 q^2 \, ,
\ee
where
\be\label{mdef}
m^2 = \left(\Delta_1^0\right)^2-\left(\Delta_2^0\right)^2+ h_2^2 \, ,
\ee
\be\label{vdef}
v_0^2 = \frac{v_F^2}{3}\left(\frac{\Delta_2^0}{\Delta_1^0}\right)^2 f\left(\frac{h_2}{\Delta_1^0}\right) \, ,
\ee
and
\be
f(x)=\frac{1-x^2}{(1+x^2)^2} \, .
\ee
There are two branches with positive, $\omega>0$, and negative, $\omega<0$, energies. Similarly to the
case of a polarized normal two-component gas \cite{igor}, we can identify these excitations as
bifermions and biholes.
We note that the mass of these modes
explicitly depends on symmetry breaking due to a difference in coupling constants [first two terms
on the right hand side of \eref{mdef}]
or   chemical potentials. In the U$(2)$-symmetric case ($h_1=h_2=0$ and $g_1=g_2$,
so $\Delta_1^0=\Delta_2^0$), the mass vanishes due to particle-hole symmetry.
This result is independent of the weak-coupling assumption and holds
at any coupling as long as particle-hole symmetry is present.
Moreover, in the symmetric case  the collective mode speed
\rref{vdef}  reduces to the   known result for the phase mode \cite{ABmode}, $v_0 = v_F/\sqrt{3}$.
In other words, in the symmetric limit
 in addition to the phase mode, there are two more
modes with the same  dispersion. This is expected in the framework
of spontaneous symmetry breaking from U$(2)$ down to U$(1)$. Due to
condensation   into the superfluid state, the system is
invariant only under rotations that change the phase of the order
parameter and not under rotations  transforming one of the
components of $\vec\Delta=(\Delta_1,\Delta_2,0)$ into the other.
Then, to the three broken generators correspond three massless
Goldstone bosons. On the other hand, in the absence of particle-hole
symmetry, the dispersion relation is modified \cite{ssb}, and
two of the massless modes split into a massless mode with quadratic
dispersion relation and a massive one \cite{note2}.

\subsection{Collective modes at finite temperatures}

Let us consider the vicinity of the second-order phase transition N-S$_1$, so that $\Delta_1\to 0$.
In this case, the inverse propagator has a form similar to the quadratic term in the Ginzburg-Landau
expansion \re{GLe} to which it reduces in the static limit $\omega \to 0$. For $\omega\ne0$ the only
difference is that the coefficients $\alpha_i$, $\beta_i$, and $\beta_{12}$ depend on $\omega$.  The frequency
dependence of $\beta_2$ and  $\beta_{12}$ can be neglected  since they multiply  small quantities $q^2$ and
$|\Delta_1|^2$, respectively. Using \eref{effac}, we obtain
\be\label{prGL}\begin{split}
&\left[\nu D(\omega,q)\right]^{-1} = \\
& \ln\left(\frac{T}{T_{c_2}}\right) +
\frac{\beta_2}{2}\frac{v_F^2q^2}{3} +\beta_{12}|\Delta_1|^2  - \Psi\left(\frac{1}{2}\right)+
\\ &
\frac{1}{2}\left[\Psi\left(\frac{1}{2}+\frac{-\mathrm{i}(\omega+h_2)}{4\pi T}\right)+
\Psi\left(\frac{1}{2}+\frac{-\mathrm{i}(\omega-h_2)}{4\pi T}\right)\right].
\end{split}\ee
The general structure of this propagator is the standard one for superconducting fluctuations
\cite{scf}, with overdamped fluctuations typical of the time-dependent Ginzburg-Landau approach.
What is peculiar here is that the mass term is proportional to $|\Delta_1|^2$.
This makes the decay of
fluctuations in the 1-3 channel (i.e., towards superfluid S$_2$) faster than those toward the normal state. Nonetheless, they
play an important role in causing deviations from mean-field theory. To show   this, we
employ   the Ginzburg-Levanyuk criterion \cite{LL5}  for the simple case $h_1=h_2=0$ and $T \lesssim T_{c_1}$.

As is well known in the theory of second-order phase transitions, fluctuations  strongly modify the mean-field
behavior at temperatures close to the critical
one \cite{LL5}. The temperature window around the critical temperature where the fluctuations
dominate can be characterized by  the Ginzburg-Levanyuk number $Gi$, so that for
$\varepsilon \equiv |T-T_c|/T_c\gg Gi$
fluctuations are small. In three dimensions, due to amplitude fluctuations,
$Gi \propto (T_c/E_F)^4$. This result can be obtained by writing the Ginzburg-Levanyuk criterion
as \cite{LL5}
\be
\frac{T_c \chi}{\xi^3} \ll |\Delta|^2 \, ,
\ee
where
\be\label{chidef}
\chi \equiv D(0,0) \propto 1/\nu\varepsilon
\ee
 is the pair susceptibility and
\be\label{cldef}
\xi^2 \equiv [D\partial^2 D^{-1}/\partial q^2](0,0) \propto v_F^2/T_c^2\varepsilon
\ee
is the coherence length squared.  In both equations above the last term on the right is due to
fluctuations of the order parameter $\Delta_1$ itself, whose propagator has the form
similar to \eref{prGL} up to the replacement of indices $2\to 1$, but without the term
$\beta_{12}|\Delta_2|^2$; see \cite{scf}.
Using $\nu \propto m^{3/2} E_F^{1/2}$ and $\Delta \propto T_c \sqrt{\varepsilon}$,
we obtain $Gi\propto (T_c/E_F)^4$.

In the present case, we can use the same approach.
Substituting the value of the order parameter $\Delta_1$,
\eref{min1}, into \eref{prGL} and using the definitions in \esref{chidef} and \rref{cldef}, we derive for
the susceptibility and the coherence length
\be
\chi \propto \left(\nu \ln \frac{T_{c_1}}{T_{c_2}}\right)^{-1} \, , \quad
\xi \propto v_F \left(T_{c_1}\sqrt{\ln \frac{T_{c_1}}{T_{c_2}}}\right)^{-1}.
\ee
Using these expressions, we obtain for the three component case
\be\label{newgi}
Gi \propto \left(\frac{T_{c_1}}{E_F}\right)^2 \sqrt{\ln \frac{T_{c_1}}{T_{c_2}}} \, .
\ee
We see that the fluctuations in the uncondensed (1-3) channel
shrink the region of applicability of mean-field theory as soon as
$\ln T_{c_1}/T_{c_2} \gg (T_{c_1}/E_F)^4$ -- i.e., even for very small differences
in the critical temperatures.

In the context of the BCS-BEC crossover, we recall that
as the strength of the interaction grows, the ratio $T_c/E_F$ grows too. This
signals the breakdown of the mean-field approximation as the unitary limit
is approached from the BCS side. The above estimate \eref{newgi} for
the Ginzburg-Levanyuk number indicates that this breakdown happens much sooner
in the presence of a third interacting component.

\section{Summary and open problems}
\label{summ}

In this paper, we considered a  three-component (species) Fermi gas
with attractive interactions between fermionic species in the weak
coupling regime. We confirmed that there are four possible
homogeneous phases: the normal state (N) and superfluids S$_i$ for
$i=1,2$, and 3 where species $j\ne i$ and $k\ne i,j$ are paired. For
simplicity, for most of the paper we restricted our analysis to the case when the
components 1 and 2 do not interact with each other. In this case,
the homogeneous phases of the system are N, S$_1$ (2 and 3 are
paired), and S$_2$ (1 and 3 are paired). The extension of  our
findings to the general case of nonzero interaction between all
components is straightforward; see Fig.~\ref{fig:g3} and Sec.~\ref{g3em}.

We   constructed the ``high''-temperature $T>T_{\mathrm{tri}}$ [see Eq.~\re{triem}] and
zero-temperature phase diagrams for arbitrary differences between
chemical potentials of the three species (Figs.~\ref{fig:1} and
\ref{fig:2}). In particular, we identified the regions where
different superfluid states and the normal state are (meta)stable
and determined the lines of first-order S$_1$-S$_2$ and second-order
N-S$_1$ and N-S$_2$  phase transitions. We also obtained
 the phase diagram in the canonical ensemble in the space of particle density
differences $(n_3-n_2)$ and $(n_1-n_3)$ (Fig.~\ref{fig:1b}).  This phase diagram displays regions where the uniform superfluid
states are unstable.  Phase separation between superfluids S$_1$ and S$_2$ occurs for particle
densities within these regions; i.e., the system becomes spatially inhomogeneous.

We analyzed the properties of the domain walls between
superfluid states S$_1$ and S$_2$. The domain walls are present in
the phase-separated region and at an interface between layers of
S$_1$ and S$_2$ in a trapped three-component gas; see the text below
Eq.~\re{locapp}. We  determined the shape of the domain wall [see
Fig.~\ref{fig:3} and Eq.~\re{shapeem}] and demonstrated that its
thickness $\ell$, \eref{ldef}, provides a new length scale  that can be
parametrically larger than the coherence lengths $\xi_{1,2}$ of
superfluids S$_{1,2}$, $\ell\gg\sqrt{\xi_{1}^2+\xi_{2}^2}$. In
particular, this means that the two order parameters of superfluids
S$_1$ and S$_2$ can overlap significantly over extended regions of
space. It also imposes severe restrictions on the LDA
for evaluating the configuration of superfluid and
normal layers in a trap\cite{review}; see the discussion  below
Eq.~\re{locapp} and in the end of Sec.~\ref{sec:var}. The sharp boundaries between the superfluids predicted by
the LDA have to be smeared over the length scale $\ell$ (rather than   the coherence lengths $\xi_{1}$ or $\xi_{2}$
).  Furthermore, the LDA is valid only when the size
of the trap, $R$,  is much larger than the domain wall thickness,
 $R\gg \ell$. Otherwise, the two superfluids coexist throughout the trap.
 For experimentally attainable systems, the condition  $R\gg \ell$ translates into the total number of atoms  $N_t\gg 10^4$ with
 corrections to the LDA being significant even for  typical numbers   in experiments, $N_t\sim 10^7$;
 see \eref{Nest} and the text after it. We also evaluated the surface tension associated with the domain wall,
Eqs.~(\ref{stapp}) and (\ref{stvar}), which needs to be taken into account when  considering the shape of the interface between
superfluids S$_1$ and S$_2$.

Finally, we studied the collective modes (fluctuations) specific to
our system in Sec.~\ref{sec:cm}. Namely, in the superfluid state S$_1$ with
order parameter $\Delta_1$ there are fluctuations $\delta
\Delta_2(\rr, t)$ of the order parameter of superfluid  S$_2$ and
vice versa. We evaluated the mass and the dispersion relations of
these collective modes at zero temperature and in the vicinity of
the N-S$_1$ transition. At $T=0$ the mass is determined by
perturbations that break the U$(2)$ symmetry between species 1 and
2 -- the difference in chemical potentials and coupling constants
for the interaction with 3. In the symmetric case the mass vanishes.
At small symmetry breaking the collective modes soften and their
mass can be parametrically smaller than the BCS energy gaps of
superfluids S$_1$ and S$_2$.
 Similarly, near the critical temperature of the N-S$_1$ these fluctuations can significantly
increase the Ginzburg-Levanyuk number [see \eref{newgi}] in comparison to the two-component system.
This indicates that stronger deviations from mean-field theory are possible in a three-component system.

The results outlined above were obtained in the weak-coupling BCS limit. A natural question is
how they are modified in the BCS-BEC crossover regime and in particular at the unitary limit for
two of the three components
when the corresponding scattering length diverges. In the two-component
case, there is a single length and energy scale at unitarity at $T=0$. This is not so in our case if the symmetry between the components
is broken. Therefore, we expect qualitatively the same picture such as extended domain walls, soft modes,
etc.,  as long as no true bound states are formed.
It is also interesting to study these phenomena at lower temperatures close
and below the tricritical temperature \re{hight} which limits the applicability of our
Ginzburg-Landau approach.

Let us also emphasize that   to make more quantitative predictions about the possible
experimental realization and detection of coexisting multiple superfluid states,  it is necessary to go
beyond or at least improve  the LDA. Further work is
also required to understand the effects of fluctuations in the unpaired channel on experimentally accessible quantities
such as  critical temperatures and the frequencies of collective oscillations in trapped gases
 in the hydrodynamic regime \cite{hydro}.

\acknowledgments

We thank B. L. Altshuler for useful discussions. This work was
financially supported by NSF award Grant No. NSF-DMR-0547769. E.A.Y. also
acknowledges the financial support of the David and Lucille Packard
Foundation and the Alfred P.
Sloan Research Foundation.

\appendix

\section*{Appendix}

In this appendix we present the expressions for the functions $H$ and $J$ introduced
in \eref{prop} for the fluctuation propagator. The method to derive these functions
is explained in \cite{ssb}. So here we limit ourselves to
the final results, which are straightforward extensions of those found
in \cite{ssb}:
\begin{eqnarray}
&& H=\frac{1}{2\nu}\int\!\frac{d^3p}{(2\pi)^3} \, \frac{1}{2E_p} \label{heq}\\
&&\bigg[\frac{1-f(E_p-h_1/2)-f(\xi_{p-q} -h_1/2+h_2)}{\omega - E_p-\xi_{p-q}+h_1-h_2}
- \nonumber \\ && \frac{f(E_p+h_1/2)-f(\xi_{p-q}-h_1/2+h_2)}{\omega + E_p - \xi_{p-q}+h_1-h_2}
+ (q \to -q)\bigg], \nonumber
\end{eqnarray}

\begin{eqnarray}
&&J=\frac{1}{2\nu}\int\!\frac{d^3p}{(2\pi)^3} \, \frac{\xi_p -\xi_{p-q}}{2E_p} \label{jeq}\\
&& \bigg[\frac{1-f(E_p-h_1/2)-f(\xi_{p-q} -h_1/2+h_2)}{\omega - E_p-\xi_{p-q}+h_1-h_2}
- \nonumber \\ && \frac{f(E_p+h_1/2)-f(\xi_{p-q}-h_1/2+h_2)}{\omega + E_p - \xi_{p-q}+h_1-h_2}
\bigg] + (q \to -q), \nonumber
\end{eqnarray}
where
\be
\xi_p = \frac{p^2}{2m}- \frac{\mu_2+\mu_3}{2} \, , \quad
E_p = \sqrt{\xi_p^2 + \Delta_1^2} \, .
\ee
In the limit $h_i \to 0$, \esref{heq} and \rref{jeq} reduce (up to a normalization factor) to the functions
$H$ and $J$ obtained in \cite{ssb}.

  We note that in deriving, e.g., \eref{prop0T} we linearize the spectrum near
the Fermi surface and assume particle-hole symmetry. Namely, we
parametrize the momentum as $\boldsymbol{p} = \nn (p_F +\xi/v_F)$, where $p_F$ is the Fermi momentum, $v_F$
the Fermi velocity, and $\nn$ the unit vector on the Fermi sphere. Then, the integral over momentum is replaced with
the integral over  $\xi$ and the vector $\nn$
\be
\int\!\frac{d^3 p}{(2\pi)^2} \to \nu \int\!d\xi \int \frac{d\nn}{4\pi} \, ,
\ee
where $\nu$ is the density of states at the Fermi energy. Going beyond this approximation would enable
the study of particle-hole asymmetry effects.


\begin{thebibliography}{99}
\bibitem{note1} This approximation is valid when the chemical potential differences
$h_i$ are of the order of the critical ones (i.e., we are in the vicinity of the second-order
phase transitions), and
the temperature is not too close to the critical one, $\ln (T_{c_2}/T) \gg (T_{c_2}/\mu_0)^2$.
\bibitem{noteu}This limit is not physical because if $a=0$, then $\alpha_1=\alpha_2$, and by
\eref{degcon}, $\beta_1=\beta_2$. This requires $h_1=\pm h_2$, which is compatible
with $\alpha_1=\alpha_2$ only if $T_{c_1}=T_{c_2}$, while we assume $T_{c_1}>T_{c_2}$.
\bibitem{notel}For simplicity, we use the same length $\ell_\mathrm{v}$ for both order parameters, which
makes an analytical treatment possible. If the coherence lengths [see \eref{xitil}] are vastly different
($|\tilde{\xi}_1-\tilde{\xi}_2|\gg \tilde{\xi_1}+\tilde{\xi_2}$), a better choice for the variational
functions would employ two separate scales. However, for the estimates at the end of this section,
the lower bound on the length scale found with this simpler treatment is sufficient.
\bibitem{note1b} For the estimates we use the simpler high-temperature formulas
of Sec.~\ref{sec:dwtc}. The results are unchanged if the variational results are used instead.
\bibitem{FFLOnote}We stress that, because of the ``high-temperature'' assumption,
the FFLO state is not relevant to our study of the spatial variation of the order
parameter due to a domain wall.
\bibitem{note2}We note that the dispersion relations presented in Ref.~\cite{ssb}
are incorrect, as they predict for a collective mode the
same dispersion relation as for free fermions; for positive and large chemical potentials, on the
other hand, the mode velocity should be
proportional to the Fermi velocity, as we show in the main text.
\end{thebibliography}
\end{document}